\documentclass[aps,preprint,showpacs,preprintnumbers,amsmath,amssymb,floatfix]{revtex4}
\usepackage{graphicx}

\begin{document}

\title{Role of Hyperons in Neutron Stars} 

\author{J. R. Stone}
\affiliation{Department of Physics, University of Oxford, OX1 3PU Oxford, 
United Kingdom}
\affiliation{Department of Physics and Astronomy, University of Tennessee, 
Knoxville, Tennessee 37996, USA}
\author{P. A. M. Guichon}
\affiliation{SPhN/IRFU, CEA Saclay, F91191 Gif-sur-Yvette Cedex, France }
\author{A. W. Thomas}
\affiliation{CSSM, School of Chemistry and Physics, University of Adelaide, Adelaide SA 5005 Australia \\}
\email{j.stone@physics.ox.ac.uk; pierre.guichon@cea.fr; Anthony.Thomas@adelaide.edu.au}

\date{\today}

\begin{abstract}
The latest observation of a Shapiro delay of the binary millisecond pulsar J1614-2230 by Demorest et al. \cite{demorest2010} yielded the pulsar gravitational mass to be 1.97$\pm$0.04 M$_\odot$, the heaviest observed pulsar to-date. This result produces a stringent constraint on Equation(s) of State (EoS) of high density neutron star matter. One of the main conclusions of Demorest et al. was that their result makes the presence of non-nucleonic components in the neutron star matter unlikely. We compare the result with our recent work and conclude that hyperons in high-density matter are fully consistent with the observation and that their presence is a necessary consequence of general physical laws. 
\end{abstract}
\pacs {97.60.Jd, 26.60.Dd, 26.60.Kp, 26.60.-c, 14.20.Jn}
\maketitle
From a strong Shapiro delay signature seen in radio timing observations of the binary millisecond pulsar J1614-2230, Demorest et al. \cite{demorest2010} calculated the pulsar gravitational mass to be 1.97$\pm$0.04 M$_\odot$, the heaviest observed pulsar to-date. In addition, they inferred an upper limit on the maximum density of observable cold matter of 3.74$\pm$0.15 x 10$^{\rm 15}$ gcm$^{\rm -3}$ ($\sim$10 times nuclear saturation density, n$_{\rm s}$) from their mass measurements. They correctly concluded that their result significantly constrains the cold nuclear matter EoS which forms input for the Tolman-Oppenheimer-Volkoff (TOV) equation, yielding a mass-radius relation for cold neutron stars. However, based on only a limited selection of `typical' EoS they come to the conclusion that their result  makes non-nucleonic components in the neutron star interior unlikely. On the contrary, by comparing the observational results of Demorest et al. \cite{demorest2010} with our recent work \cite{stone2007}, we conclude that hyperons in high-density matter are completely consistent with the observation.

Cold neutron stars with masses in the range (1.99 - 1.90) M$_\odot$ were predicted in the framework of the Quark-Meson-Coupling model (QMC) three years ago, assuming matter containing the full baryon octet (nucleons and hyperons). The corresponding radii, central densities and central pressures are (12.45 - 11.93) km and (1.66 - 1.74) x 10$^{\rm 15}$ gcm$^{\rm -3}$, [(6.61 - 6.93) n$_{\rm s}$] and (182 - 209) MeVfm$^{\rm -3}$. The gravitational red shift is 0.38, in line with observational constraint \cite{lackey2006}, and the speed of sound is 0.66 c. We note that the central densities are well below the upper bound quoted by Demorest et al. and that the radii of the stars, compatible with the new observation, are clustered very close to 12.35 km.

\begin{figure}[ht]
\begin{center}
\includegraphics[scale=0.6]{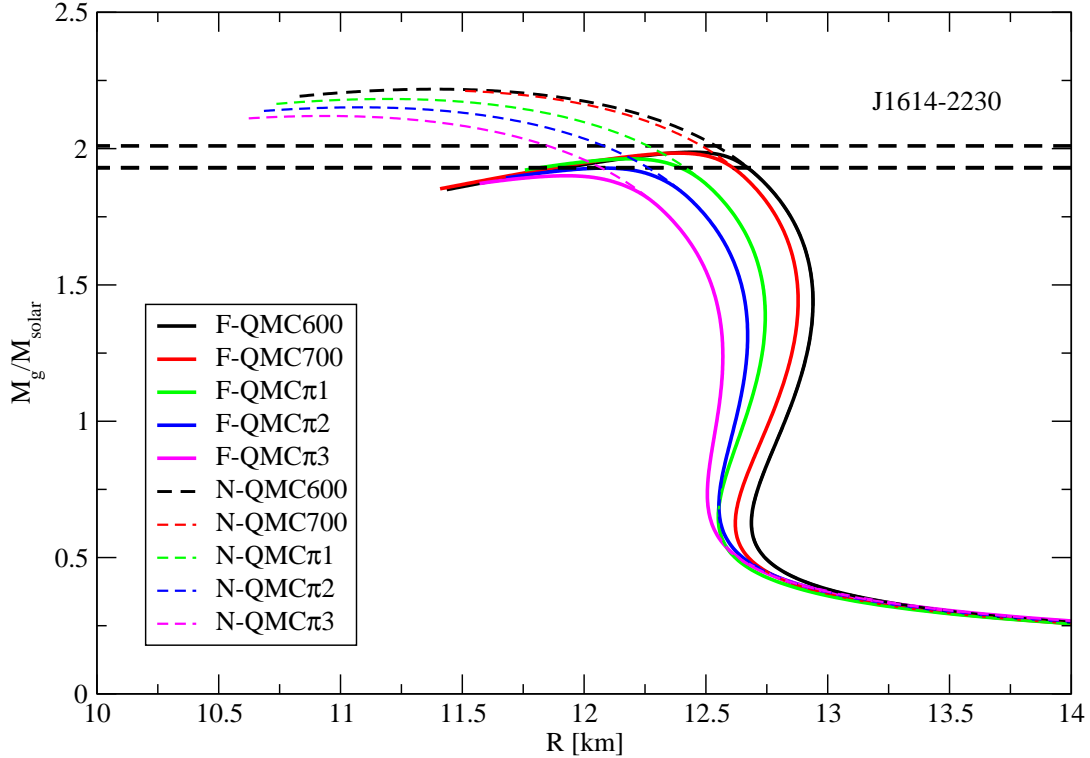}
\caption{(color online)\label{fig1} The gravitational mass of cold non-rotating neutron star models plotted against radius as calculated in the QMC model - see text for more explanation.}
\end{center}
\end{figure}

In Fig.~\ref{fig1} we show the mass-radius curves for five (closely related) versions of the QMC model, which include the full baryon octet (F-QMCx) and five more (again closely related), which have nucleons only (N-QMCx) (for details see \cite{stone2007}). The presence of hyperons in the neutron star matter reduces the maximum mass because of the softening of the EoS above a certain central density as compared to the prediction for nucleon-only matter. The magnitude of this effect depends on the model used for the nucleon-nucleon and nucleon-hyperon interactions.  Fig.~\ref{fig1} illustrates that the QMC model does indeed predict neutron stars consistent with the new observation when the full baryon octet is included.   

The QMC model takes into account the modification of the quark and gluon structure of the baryons in the medium and may therefore differ from other models in significant ways when strangeness appears. The uncertainty in the hyperon-nucleon coupling constants, inherent to relativistic field models \cite{lackey2006}, is not present in this model. It correctly reproduces the binding energy of the ground states of known $\Lambda$ hypernuclei while not binding $\Sigma$ hypernuclei \cite{guichon2008}. Furthermore, it should be stressed that, whilst derived at the quark level, the QMC model generates very realistic Skyrme forces \cite{guichon2006,dutra2010} for modelling of non-equilibrium matter and finite nuclei.

The argument that hyperons should play a role in the structure of neutron stars is quite compelling. Because of the Pauli Principle, the chemical potential for neutrons rises rapidly with increasing density, reaching that for the $\Lambda$ (and within QMC the $\Xi_0$) hyperons at (2 - 3) n$_{\rm s}$. It then becomes energetically favourable for the system to let the neutrons undergo a strangeness changing weak decay, which replaces them by hyperons, for which the Fermi sea is not yet filled, thus lowering the total energy of the system.  As strangeness is not conserved on the weak interaction scale and the time-scales for neutron star formation are much greater than those associated with weak interactions, the grow of strangeness  will continue until equilibrium is reached. This means that any hyperon energetically allowed \textit{ must} appear. Rather than being a surprise to find hyperons it would stretch our understanding of fundamental strong and weak interaction processes to breaking point if they were not to appear. It is certainly inconceivable that a nucleon-only EoS could be realistic at such large densities. 

We conclude that the Demorest et al. result, if confirmed, is very significant for neutron star physics and does indeed rule out all EoS, which predict a mass-radius curve that does not intersect the J1614-2230 mass line. However, the fact that almost all \textit{currently available} EoS including hyperons are ruled out, does not imply the non-existence of hyperons in the high-density matter typical of such a massive neutron star. Rather it poses a challenge to theorists to pay more attention to the development of EoS, such as that derived within QMC, which describe the properties of observed hypernuclei and therefore provide a reasonable basis for describing the composition of high-density neutron star matter consistent with general laws of physics as well as with the latest observations.

\end{document}